\documentclass{aastex62}
\usepackage{graphicx}	
\usepackage{amsmath}	
\usepackage{amssymb}	
\usepackage{natbib}     
\usepackage{subfigure}
\usepackage{booktabs}
\usepackage{tabularx}
\usepackage{mathrsfs}

	\graphicspath{{./}{figures/}}

	\received{}
	\revised{}
	\accepted{}
	\submitjournal{ApJ}
	
\begin{document}
			
			\title{
				Rapid Rotation of Polarization Orientations in PSR B1919$+$21's Single Pulses: \\Implications on Pulsar's Magnetospheric Dynamics
			}%

			\correspondingauthor{Jinchen Jiang, Kejia Lee, Renxin Xu}
			\email{jiangjinchen@bao.ac.cn, kjlee@pku.edu.cn, r.x.xu@pku.edu.cn}
			
			\author{Shunshun Cao}
            \affiliation{State Key Laboratory of Nuclear Physics and Technology, Peking University, Beijing 100871, China}
            \affiliation{Department of Astronomy, School of Physics,
				Peking University, Beijing 100871, China}
			
			\author{Jinchen Jiang}
			\affiliation{National Astronomical Observatories, Chinese Academy of Sciences, Beijing 100012, China}
			
			\author{Jaroslaw Dyks}
			\affiliation{Nicolaus Copernicus Astronomical Center, Polish Academy of Sciences, Rabia\'{n}ska 8, 87-100, Toru\'{n}, Poland}
			
			\author{Kejia Lee}
			\affiliation{Kavli Institute for Astronomy and
				Astrophysics, Peking University, Beijing 100871, China}
            \affiliation{Department of Astronomy, School of Physics,
				Peking University, Beijing 100871, China}
			\affiliation{National Astronomical Observatories, Chinese Academy of Sciences, Beijing 100012, China}
			
			\author{Jiguang Lu}
			\affiliation{National Astronomical Observatories, Chinese Academy of Sciences, Beijing 100012, China}
			\affiliation{Guizhou Radio Astronomical Observatory, Guiyang 550025, China}

            \author{Lucy S. Oswald}
			\affiliation{School of Physics and Astronomy, University of Southampton, Southampton SO17 1BJ, UK}
			
			\author{Weiyang Wang}
			\affiliation{School of Astronomy and Space Science, University of Chinese Academy of Sciences, Beijing 100049, China}
			
			\author{Renxin Xu}
			\affiliation{State Key Laboratory of Nuclear Physics and Technology, Peking University, Beijing 100871, China}
            \affiliation{Department of Astronomy, School of Physics,
				Peking University, Beijing 100871, China}
			\affiliation{Kavli Institute for Astronomy and
				Astrophysics, Peking University, Beijing 100871, China}

			
			
			\begin{abstract}
				
                We analyze and model rapid rotations of polarization orientations in PSR B1919$+$21's single pulses based on Five-hundred-meter Aperture Spherical radio Telescope observation data. In more than one-third of B1919$+$21's single pulses, the polarization position angle (PA) is found to rotate quasi-monotonically with pulse longitude, by over 180$^{\circ}$ or even 360$^{\circ}$. Some single pulse PA even rotates by over 540$^{\circ}$. Most of these quasi-monotonic PA curves have negative slopes with respect to pulse longitude. Oscillations of circular polarization fraction accompany these PA rotations. This rapid rotation could be induced by a quick change of phase lag between two normal wave modes within an individual pulse. We propose a phenomenological model to reproduce the observed polarization rotations in single pulses, and calculate phase lags in a dipolar magnetic field of an aligned rotating pulsar, with a dispersion relation of orthogonal wave modes in strongly magnetized electron-positron plasma. According to the dispersion relation, the weak frequency dependence of observed polarization rotation requires small angles between the radio wavevector and local magnetic fields, which requires the radio emission height to be low, on the order of 10 times neutron star radius.
			\end{abstract}
			
		
		
		\section{Introduction} \label{sec:intro}
		Magnetospheres around neutron stars generate and modify pulsar radio emission (see \citealp{2022ARA&A..60..495P} for a review). Magnetospheres not only play a vital role in pulsar and neutron star theories, but are also natural laboratories for particle physics~\citep[e.g.,][]{2023PhRvL.131k1004N, 2023PhRvD.108h3009X}.

        Among theoretical studies, especially analytical studies, much effort has been focused on analyzing normal wave modes in magnetospheres. A fundamental property of the magnetosphere is the birefringence, so dispersion relations of the ordinary (O) mode and the extraordinary (X/E) mode are the basis for further analysis of wave propagation. \cite{1977PASA....3..120M} derived the dispersion relations of O and X modes in a cold relativistic plasma with a single bulk velocity value and infinitely large magnetic field strength, and naturally explained pulsars' orthogonal polarization modes observed by~\cite{1975ApJ...196...83M}. Such dispersion relations have been applied to describe ray trajectories in pulsar magnetospheres~\citep[e.g.,][]{1986ApJ...302..138B, 2000A&A...355.1168P} and to explain the origin of circular polarization in pulsar radio emissions~\citep[e.g.,][]{1998Ap&SS.262..379L}, involving calculating the phase lag between wave modes. 
        
        Despite the success of models, some parameters in pulsar magnetospheres are not well constrained. Two key parameters are the plasma multiplicity $\kappa$, defined as the ratio between the actual plasma density and the Goldreich\text{-}Julian particle number density (Equation (9) in \citealp{1969ApJ...157..869G}), and the Lorentz factor of the secondary plasma particles, $\gamma_{\mathrm{sec}}$. The secondary plasma particles are electron positron pairs produced in polar cap pair cascade processes. The Lorentz factor of these particles is also related to the emission height, the distance between the emission point, and the pulsar centroid. Usually $\kappa$ takes the value $10^{3}\sim 10^{5}$, and $\gamma_{\mathrm{sec}}$ takes the value $10\sim 10^{3}$~\citep[e.g.,][]{2022ARA&A..60..495P}. Beside some attempts to achieve more realistic simulations, studying observations' connection with theories is important for pushing studies forward. The double pulsar system J0737$-$3039 has provided good observational constraints on the magnetospheric physics~\citep{2005ApJ...634.1223L,2024MNRAS.534.3936L}. However, for solitary pulsars, detailed studies of individual pulse behaviors might bring us more insight into pulsar magnetospheres, with China's Five-hundred-meter Aperture Spherical radio Telescope (FAST) now providing a large amount of high\text{-}quality radio pulsar data.

        In this paper, we present our analysis and modeling of some rapid polarization position angle (PA) rotations in PSR~B1919$+$21's single pulses, based on FAST data. As the first pulsar ever discovered~\citep{1968Natur.217..709H}, B1919$+$21 is frequently studied under various frequency bands. It is a typical normal pulsar with spin period $P=1.34$~s and period derivative $\dot{P}=1.35\times 10^{-15}$~\citep{2005AJ....129.1993M}. B1919$+$21 exhibits sub-pulse drifting, with drifting properties changing with frequency~\citep{1975ApJ...195..193C, 1986ApJ...307..540P}. It was also found to exhibit unusual radius\text{-}frequency mapping (RFM), with its pulse profile width slightly increasing as the frequency increases~\citep{2002ApJ...577..322M, 2012A&A...543A..66H, 2016A&A...586A..92P}. Besides, polarization studies on this pulsar have shown that the PA distribution of single pulses changes significantly with frequency~\citep{2015ApJ...806..236M, 2019MNRAS.489.1543O}. \cite{2017MNRAS.472.4598D, 2019MNRAS.488.2018D} modeled the 45$^{\circ}$ PA jumps observed in B1919+21~\citep{2015ApJ...806..236M} through the coherent superposition of orthogonally polarized proper-mode waves.~\cite{2022A&A...657A..34P} have identified a slow rotation of polarization direction within a limited longitude range. The rotations occurred on the time scale of subpulse drift and led to a torus-shaped distribution of polarization states in Stokes parameter space ($Q/I,U/I,V/I$). Our work will further investigate single pulses' polarization patterns, and try to relate them to theoretical issues in pulsar magnetospheres.

        In Section~\ref{sec:obs}, basic information on observation and data reduction is given. Section~\ref{sec:results} shows the observed results, ranging from the basic properties of the integrated profile and the highly linearly polarized profile, to single pulses' distribution patterns, and finally, the rapid rotating polarization orientations in individual single pulses. In Section~\ref{sec:modeling} we develop the model to explain the quick rotating single\text{-}pulse polarization orientations in two steps: firstly a phenomenological model for reproducing the observed polarization patterns and exploring the parameter space, secondly the normal modes' phase lag calculation in a more realistic magnetosphere, with known dispersion relation. We propose that the key factor for understanding the polarization pattern is the phase lag value between orthogonal modes. Discussions on model feasibility, possible alternative models, and asymmetries in B1919$+$21's polarization pattern are presented in Section~\ref{sec:discussion}. Section~\ref{sec:conclusion} summarizes the conclusions.

		\section{Observation and data reduction}\label{sec:obs}
		Two epochs of observation on 2023 August 23 and 2023 September 30, are used in this paper. All of them were made by FAST at small zenith angles ($ < 26^{\circ}_{.}4$) to optimize sensitivity and polarimetry accuracy, with the $L$-band 19-beam receiver \citep{2020RAA....20...64J}. The data were recorded under the frequency band 1000--1500~MHz, and the band was divided into 4096 channels. The time resolution of the recording is 49.152$\mu$s. At the beginning of each observation, modulated signals from a noise diode were injected as a 100\% linearly polarized source for polarimetric calibration. The data processing (including folding, RFI mitigation, calibration, and timing) is done with the software packages \textsc{dspsr} \citep{2011PASA...28....1V}, \textsc{psrchive} \citep{2004PASA...21..302H}, and \textsc{tempo2} \citep{2006MNRAS.369..655H}. We get 2947 pulses in total, and each pulse period is divided into 4096 bins. In this paper, we follow the PSR/IEEE convention for the definition of Stokes parameters \citep{2010PASA...27..104V}. To correct the bias in calculating linear polarization intensity $L$ through $L=\sqrt{Q^{2}+U^{2}}$, we follow~\citet{2001ApJ...553..341E} and~\citet{2022RAA....22l4003J}\footnote{In~\citet{2022RAA....22l4003J}, the description of Equation (7) is not precise because when calculating $\epsilon_{P}$, the summation is actually made for $i,j,k=(1,2,3)$, $(2,3,1)$ and $(3,1,2)$, rather than all possible permutations.} to correct $L$ as
        \begin{align}
            & L = \left\{
            \begin{aligned}
               & \sqrt{Q^{2}+U^{2}-\epsilon_{L}},&Q^{2}+U^{2}-\epsilon_{L}>0\\
               & 0,& Q^{2}+U^{2}-\epsilon_{L}<0
            \end{aligned}
            \label{eq:L_bias}
            \right.
        \end{align}
        \noindent where
        \begin{equation}
            \epsilon_{L}=\dfrac{Q^{2}\sigma_{U}^{2}+U^{2}\sigma_{Q}^{2}}{Q^{2}+U^{2}}
            \label{eq:epsilon_L}
        \end{equation}
        Similarly, for calculating total polarization intensity $P$, we have
        \begin{align}
            &P = \left\{
            \begin{aligned}
            & \sqrt{Q^{2}+U^{2}+V^{2}-\epsilon_{P}},& Q^{2}+U^{2}+V^{2}-\epsilon_{P}>0\\
            &0,&Q^{2}+U^{2}+V^{2}-\epsilon_{P}<0
            \end{aligned}
            \right.
            \label{eq:P_bias}
        \end{align}
        where
        \begin{equation}
            \epsilon_{P}=\dfrac{Q^{2}(\sigma_{U}^{2}+\sigma_{V}^{2})+U^{2}(\sigma_{V}^{2}+\sigma_{Q}^{2})+V^{2}(\sigma_{Q}^{2}+\sigma_{U}^{2})}{Q^{2}+U^{2}+V^{2}}
            \label{eq:epsilon_P}
        \end{equation}
        Errors of $L$ and $P$ could be derived from equations above with error propagation. We skip the details here.

		\section{Results}\label{sec:results}
		
		\subsection{Overview of integrated profiles' properties}\label{sec:integrated_prof}

\begin{figure}\centering
			\centering
			\includegraphics[scale=0.55]{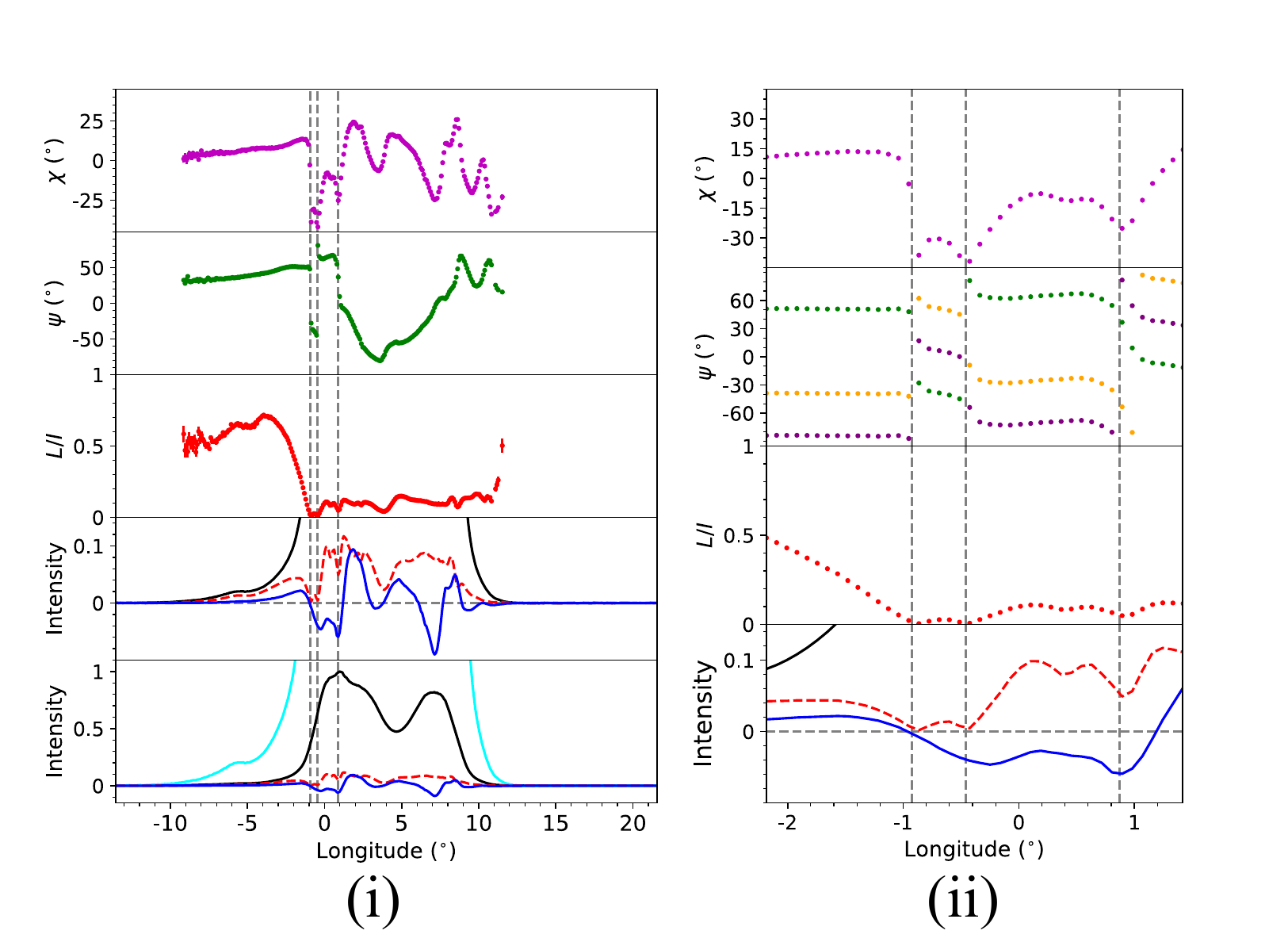}
	
		\caption{(i) Integrated profile of 2947 pulses of B1919$+$21. Black line---total intensity ($I$); Cyan line---10 times total intensity ($10\times I$); red dashed line---linear polarization intensity ($L=\sqrt{Q^{2}+U^{2}}$); blue line---circular polarization intensity ($V$). $I$, $L$ and $V$ are normalized by the maximum total intensity of the respective profiles. Red dots in $L/I$ panel with errorbars---linear polarization degrees. Green dots in $\psi$ panel with errorbars---polarization position angles (PA, $\psi=0.5\arctan(U/Q)$). Magenta dots in $\chi$ panel with errorbars---ellipticity angel (EA, $\chi=0.5\arcsin(V/{\sqrt{Q^{2}+U^{2}+V^{2}}})$). A zoomed in version of the $I$, $L$, and $V$ profiles are added as the second panel (from bottom to top). Three vertical gray dashed lines represent the longitudes $-0^{\circ}_{.}925$, $-0^{\circ}_{.}457$ and $0^{\circ}_{.}875$. (ii) A more zoomed in version of $I$, $L$, $V$ profiles, $L/I$, $\psi$ and $\chi$ panels. In the $\psi$ panel, a purple dotted curve and a yellow dotted curve are added, which represents the PA curve shifted by $45^{\circ}$ and $90^{\circ}$. Only longitudes where $L/\sigma_{L}>10$ are chosen for plotting ($\sigma_{L}$ is the standard deviation of $L$ calculated from Equation~\ref{eq:L_bias} through error propagation). \label{fig:int_prof}}
	  \end{figure}

        The integrated profile of all pulses is shown in Figure~\ref{fig:int_prof}, where the total intensity ($I$), linear polarization intensity ($L$), circular polarization intensity ($V$), PA ($\psi$), and ellipticity angle (EA, $\chi$) at each longitude are plotted. There is a precursor component on the leading edge of the profile, which contributes to the long extension of PA and EA curves to the left. Definitions of PA and EA are in Equation~\ref{eq:PA}. The error of PA could then be derived from error propagation: 
        \begin{equation}
            \sigma_{\mathrm{PA}}=\dfrac{1}{2}\dfrac{\sqrt{Q^{2}\sigma_{U}^{2}+U^{2}\sigma_{Q}^{2}}}{Q^{2}+U^{2}}
            \label{eq:sig_PA}
        \end{equation}
        The EA from the data is derived as
        \begin{equation}
            \mathrm{EA} = 0.5\arctan(V/L)\label{eq:EA}
        \end{equation}
        \noindent where $L$ is defined in Equation~\ref{eq:L_bias}. And the error of EA is similar to that of PA:
        \begin{equation}
            \sigma_{\mathrm{EA}}=\dfrac{1}{2}\dfrac{\sqrt{L^{2}\sigma_{V}^{2}+V^{2}\sigma_{L}^{2}}}{L^{2}+V^{2}}
            \label{eq:sig_EA}
        \end{equation}
        
        The polarization profiles are in accord with the results in~\citet{1999ApJS..121..171W}. B1919$+$21's polarization is complex, with a PA curve that vastly deviates from rotating vector model (RVM) curves~\citep{1969ApL.....3..225R}. Except for longitudes around the leading and trailing edges of the pulse profile, the linear polarization fraction ($L/I$) is small ($<0.2$). 
        
        In the longitude range of the leading component, the PA curve shows three ``discontinuities'', which are marked with three vertical gray dashed lines in Figure~\ref{fig:int_prof}. The first one is a nearly $90^{\circ}$ jump at longitude $-0^{\circ}_{.}925$, accompanied by a local minimum in $L$ and $L/I$, and a sign change of $V$ and $\chi$. The second one is a $45^{\circ}$ ($135^{\circ}$, most clear in Figure~\ref{fig:int_prof} (ii)) jump at longitude $-0^{\circ}_{.}457$ with a local minimum in $L$ and $L/I$, and a local maximum in $|V|$ and $|\chi|$. The third one is a nearly $45^{\circ}$ mild jump at longitude $0^{\circ}_{.}875$ with a local minimum in $L$ and $L/I$, and a local maximum in $|V|$ and $|\chi|$. After the third ``discontinuity'', the PA curve is continuous but distorted, characteristic with low $L/I$, in the central region of the pulse profile. The $V$ and $\chi$ curves are also complex. The first ``discontinuity'' could be interpreted as the transition between two orthogonal polarization modes (OPMs), whose existence has been reported by previous observations~\citep[e.g.,][]{2015ApJ...806..236M}. The third one has also been recorded and discussed before~\citep[e.g.,][]{2015ApJ...806..236M, 2017MNRAS.472.4598D}, which is associated with coherent summation of OPMs. The second one, however, seems newly discovered, and may have an origin similar to the third ``discontinuity''.

		
		\subsection{Attempts at deriving radiative geometry}\label{sec:PA_geo}

\begin{figure}\centering
			\includegraphics[scale=0.22]{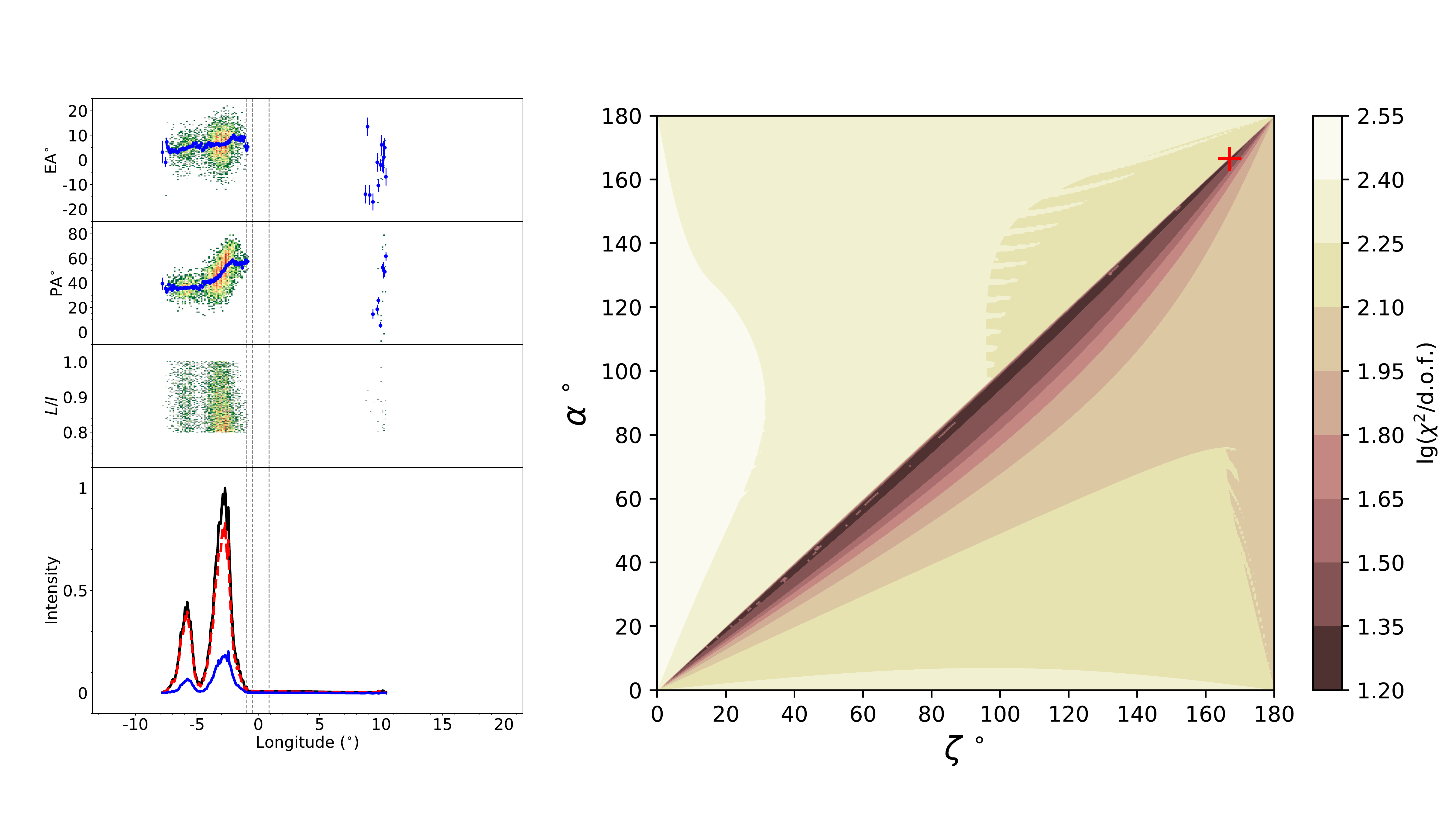}
		\caption{Left: profile from adding only highly linearly polarized pulse bins (for details, please refer to the text in Section~\ref{sec:PA_geo}). $L$ bias is also corrected here. The lowest panel: the polarization profile, the meanings of the lines are the same as those in Figure~\ref{fig:int_prof}. The other three panels: distributions of $L/I$, PA ($\psi$) and EA ($\chi$) versus longitude. The blue dots with error bars in the PA and EA panels are directly calculated (through Equations~\ref{eq:PA} and~\ref{eq:EA}) from the highly polarized pulse profile in the lowest panel. Right: the common logarithm values of chi-square (divided by degree of freedom) of fitting the RVM curve of given ($\alpha$,$\zeta$) to the modified PA curve. The red ``+'' sign marks the ($\alpha$,$\zeta$) where we get the smallest $\chi^{2}$.\label{fig:highL_prof}}
	  \end{figure}

        To derive radiative geometry, we adopt the method proposed in \cite{2023MNRAS.521L..34M} and tested in \cite{2024MNRAS.530.4839J}. Practically, at each longitude, we choose all samples having linear polarization fraction $L/I>0.8$ and PA measurement error $\sigma_{\mathrm{PA}}<5^{\circ}$, to add them up and form a highly linearly polarized profile. The result is shown in Figure~\ref{fig:highL_prof}. Compared with the total integrated profile in Figure~\ref{fig:int_prof}, highly linearly polarized signals appear only at the edges of the on-pulse window, and represent only one of the OPMs. The highly linearly polarized profile has two emission components at the leading edge of the total integrated profile. The two components are close to each other, with a large depth in between. This phenomenon is similar to the bifurcated emission component studied in~\citet{2010MNRAS.401.1781D,2023MNRAS.522.1480D}. We tried RVM curve fitting of the PA dots in the highly linearly polarized profile: for each pair of ($\alpha$,$\zeta$), we calculate the least chi-square value when changing ($\psi_{0}$, $\phi_{0}$) ($\phi_{0}$ is limited between $-90^{\circ}$ and $90^{\circ}$). The fitting process is the same as that in~\cite{2024ApJ...973...56C}. The minimum chi\text{-}squares plot is also presented in Figure~\ref{fig:highL_prof}. For B1919$+$21, the inclination angle $\alpha$ is poorly constrained, but the impact angle $\beta = \zeta - \alpha$ seems constrained to a small value, which is consistent with the estimate by~\cite{1992Ap&SS.190..209K} based on the radiation cone model. In all plots, the $\phi=0^{\circ}$ longitude is chosen as the best\text{-}fitted $\phi_{0}$, which is near the longitude of the leading component peak.

	\subsection{Single pulses' polarization distributions}\label{sec:single_pol_dist}

        \par
\begin{figure}\centering
			\includegraphics[scale=0.22]{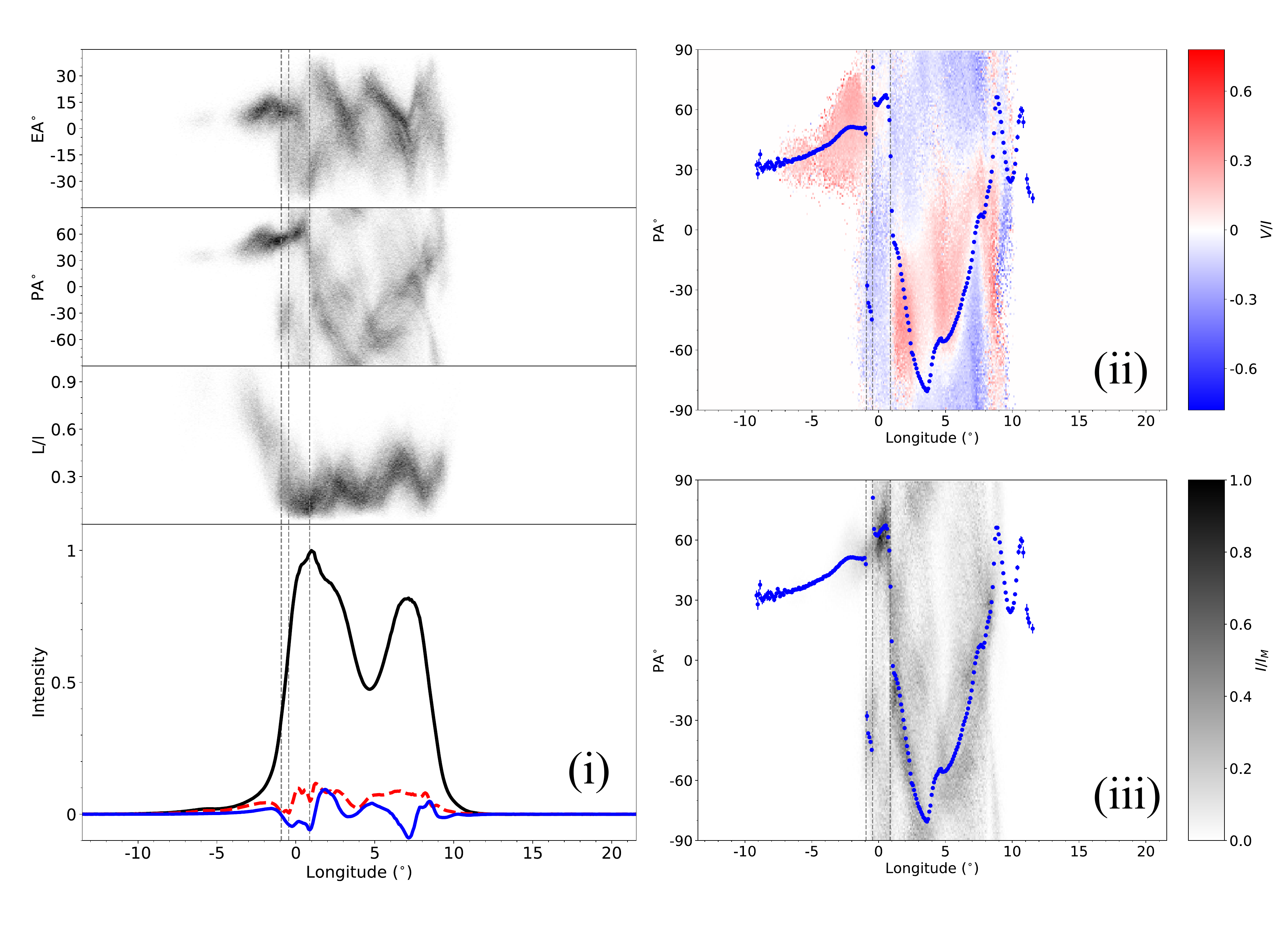}
		
		\caption{(i) Distributions of PA, EA, and $L/I$ versus longitude of all two observation epochs' 2947 pulses. Panel 4 (1\text{-}4 from top to bottom): the integrated profiles of all pulses taken into account, where the meanings of the lines are the same as Figure~\ref{fig:int_prof}; panel 3: the distributions of $L/I$; panel 2: the distributions of PA; panel 1: the distributions of EA. For panel 1, 2 and 3, the distribution is more concentrated when the color turns darker. In all panels, three vertial gray lines are the same as those in Figure~\ref{fig:int_prof}. (ii) Distribution of circular polarization fraction painted on $\phi-\psi$ plot (for calculation details, please refer to the text). (iii) Similar plot for intensity distribution (normalized with the maximum intensity value in the distribution). Only single pulses' bins where both $\sigma_{\text{PA}} \le 5^{\circ}$ and $\sigma_{\text{EA}} \le 5^{\circ}$ are included to make sure that all presented points are with significant linear and circular polarization. The blue dots with error bars are calculated from the integrated profile, the same as in Figure~\ref{fig:int_prof}.\label{fig:pulse_dist}}
	\end{figure}

 First, we calculate the 2D histogram of $L/I$, PA, and EA for all pulses in the on-pulse window, and the result is shown in Figure~\ref{fig:pulse_dist} (i). The histograms in Figure~\ref{fig:pulse_dist} (i) are made by counting the number of pulses within the longitude and PA bins. For a given longitude $\phi$ and a given PA $\psi$, we calculate the average circular polarization fraction ($\bar{V}/\bar{I}$) and the average intensity $\bar{I}$ of samples whose longitudes are within ($\phi$, $\phi+\Delta\phi$) and PAs are within ($\psi$, $\psi+\Delta\psi$). We choose $\Delta\phi=360^{\circ}/4096$, and $\Delta\psi=1^\circ$. The histograms of the average circular polarization fraction ($\bar{V}/\bar{I}$) and the average intensity $\bar{I}$ are shown in Figure~\ref{fig:pulse_dist} (ii) and (iii) respectively. In the PA panel, we directly see enhancements of single-pulse PA distributions that represent two OPMs, around the longitude range of the first two vertical gray dashed lines. On the right side of the third dashed line, even the individual pulses' PA distribution does not seem to follow RVM-like tracks, though OPMs are likely to also appear at longitude $3^{\circ} \sim 4^{\circ}$. The EA's distribution forms several tracks, varying from positive values to negative values.

    \par The $V/I$ distribution plot shows the difference in circular polarization senses for OPMs around $\phi=-1^{\circ}$, which is natural for normal wave modes in magnetized plasmas~\citep[e.g.,][]{2012MNRAS.425..814B, 2024ApJ...973...56C}. Moreover, in the central region ($1^{\circ}<\phi<8^{\circ}$), the $V/I$ sense distribution is related to the integrated profile's PA curve: in longitude range ($1^{\circ}$, $3^{\circ}$), the integrated profile's PAs correspond to $V>0$; while in longitude range ($6^{\circ}$, $8^{\circ}$), the integrated PA curve seems to be located around the $V=0$ region in the histogram for individual pulse bins. Such distributions should have some physical implications.

    \subsection{Rapid rotating polarization orientations in single pulses}\label{sec:single_pol}

\begin{figure}
		\centering
		\includegraphics[scale=0.32]{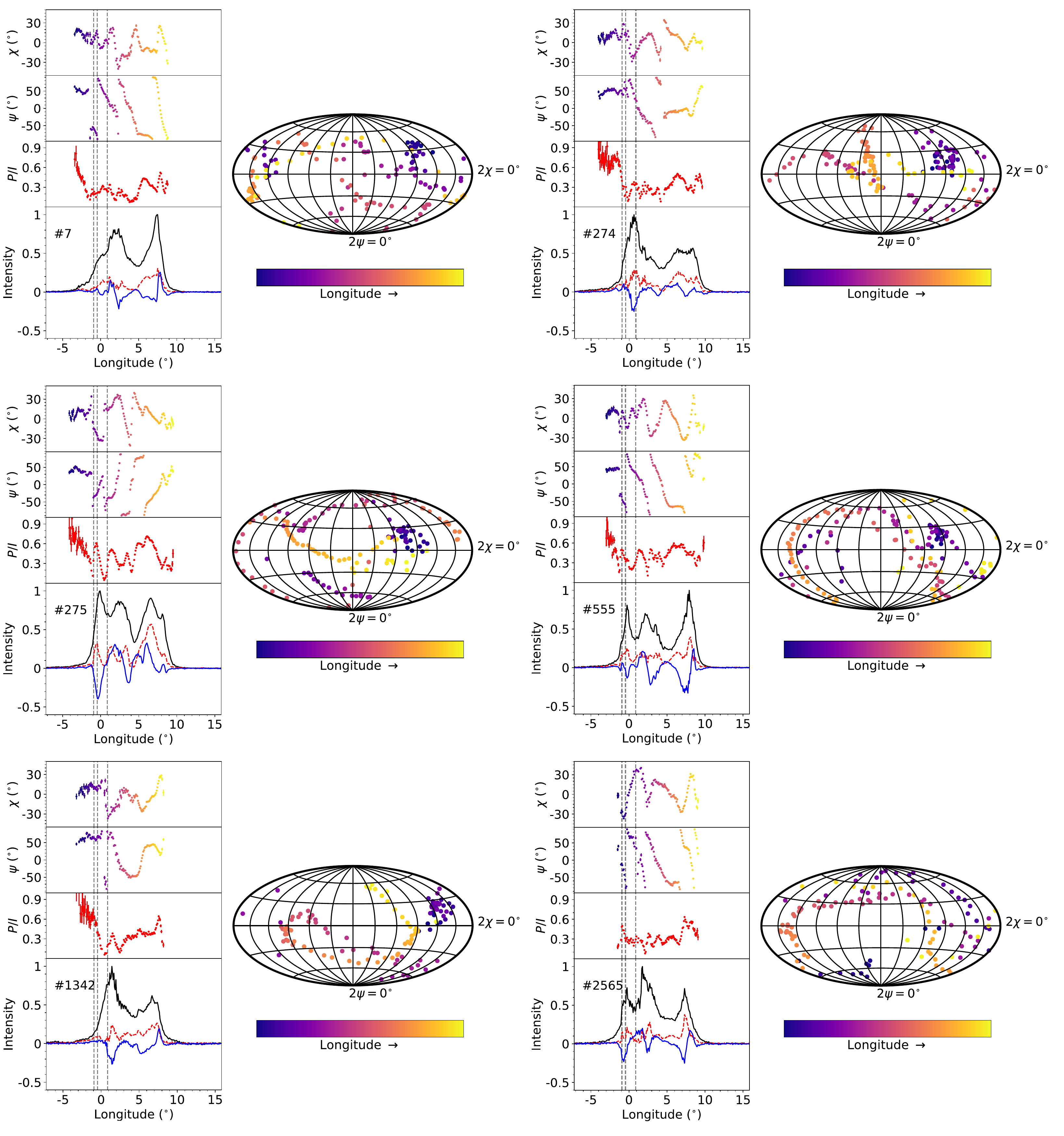}
	\caption{Six single pulses with their polarization states are shown on Poincar\'{e} spheres. In the ``Intensity'' panels, the meanings of the lines are the same as those in Figure~\ref{fig:int_prof}. ``$P/I$'' panels show the polarization fraction $\sqrt{Q^{2}+U^{2}+V^{2}}/I$ with errorbars on each longitudes (only longitudes where $\sigma_{\mathrm{PA}}\le5$ are chosen, same for the ``$\psi$'' panels and ``$\chi$'' panels). ``$\psi$'' panels show PAs with error bars, and ``$\chi$'' panels show EAs with error bars, on each chosen longitudes. PA and EA dots are also plotted on Poincar\'{e} spheres under Hammer\text{-}Aitoff projection. For PA dots, EA dots, and the track on Poincar\'{e} sphere, the change in color from darker to lighter represents the increase in longitude.\label{fig:pulses}}
    \end{figure}

    Next, we directly investigate the individual pulses. Several pulses are presented in Figure~\ref{fig:pulses}, with polarization evolution tracks on Poincar\'{e} spheres plotted alongside. The polarization states of individual pulses still seem irregular, with PA tracks appearing arbitrary. Not all pulses show ``discontinuities'' around all three vertical gray dashed lines. Orthogonal mode jumps emerge in some pulses' core longitude ranges (e.g., \#274 in Figure~\ref{fig:pulses}, at longitude $\sim 5^{\circ}$). It is worth noting that some pulses present monotonically rotating long PA curves in some longitude ranges (see \#7, \#275, \#555 and \#2565 in Figure~\ref{fig:pulses} ,for example). The PA in those pulses rotates by over $180^{\circ}$ or even $360^{\circ}$ across longitudes, accompanied by manifest oscillations of circular polarization (or ellipticity angle) values around $0$. The polarization states at these pulse longitudes seem to follow long circular tracks on Poincar\'{e} spheres.

\begin{figure}
		\centering
		\includegraphics[scale=0.35]{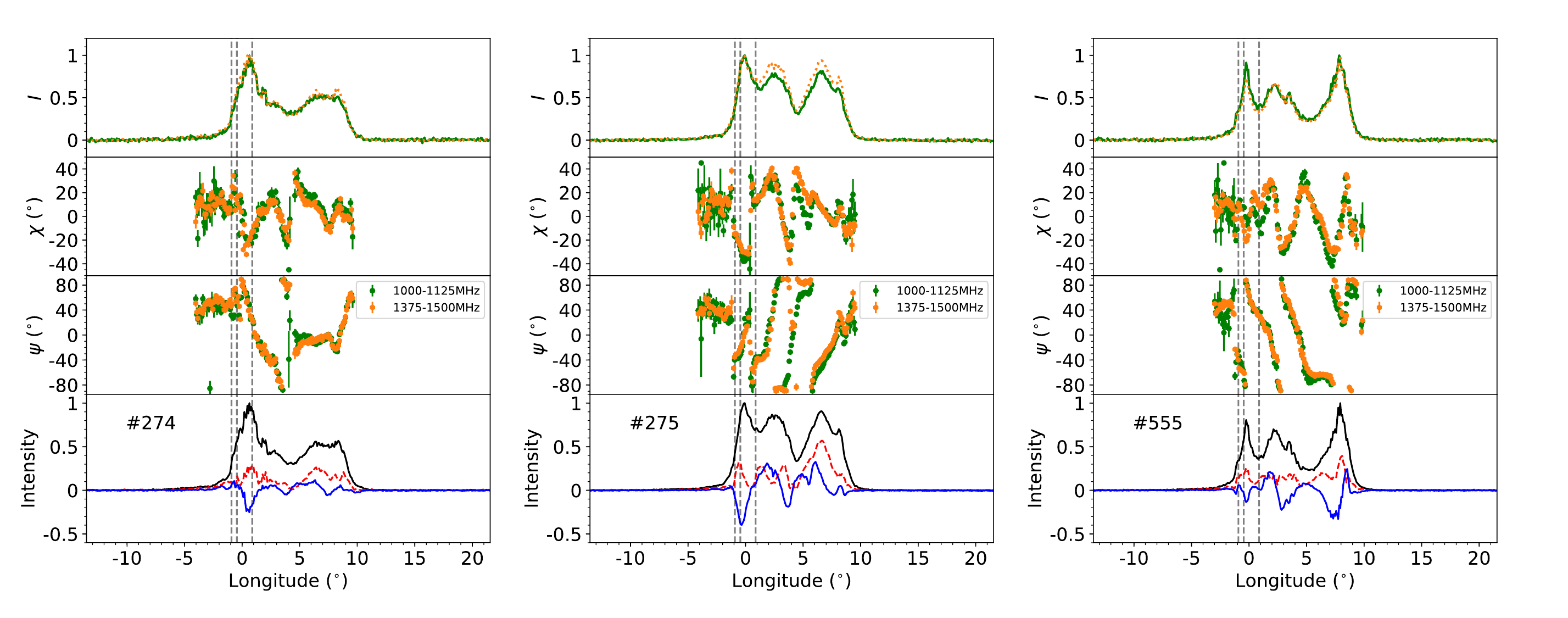}
	\caption{Three single pulses with PA ($\psi$), EA ($\chi$) and total intensity $I$ plotted in two frequency subbands (1000--1125 MHz in green and 1375--1500 MHz in orange). In the $I$ panel, the intensity curves are normalized to the maximum values separately for two subbands. The meanings of the lines and dots are the same as those in Figure~\ref{fig:pulses}.\label{fig:freq}}
    \end{figure}

    Such rapid rotating polarization patterns are not rare for B1919$+$21, with more than one-third of pulses showing quasi-monotonic rapid rotations of PA in some longitude ranges. This phenomenon seems to depend only very weakly on frequency (see Figure~\ref{fig:freq} for three examples), at least in our observational bandwidth. In addition, considering PA curves as functions of pulse longitudes, most quasi-monotonic PA curves in single pulses have a generally negative slope (e.g., \#7, \#555 and \#2565 in Figure~\ref{fig:pulses}), while only very few of them have a positive slope (e.g., \#275 in Figure~\ref{fig:pulses}). This asymmetry in PA curve slopes' distribution might have some physical implications, which we will discuss later. Finally, we would like to point out that the PA curves between two adjacent pulses (e.g., \#274 and \#275 in Figure~\ref{fig:pulses}) can change vastly.

  \section{Attempts at modeling}\label{sec:modeling}  

        Among the observed phenomena, the rapid rotating polarization orientations in single pulses are outstanding. The PA curves are far from the RVM types. Non-RVM-type PA curves have been discovered and discussed in pulsars (see, e.g.,~\citealp{2023MNRAS.525..840O}), magnetars~\citep{2024NatAs...8..606L}, and fast radio bursts~\citep{2024arXiv241114784B}. But the phenomenon in B1919$+$21 is unique in its extreme: in some pulses the PA rotates by over 540$^{\circ}$ (see \#2565 in Figure~\ref{fig:pulses} for example). OPM jumps in integrated profile and single pulses lead us to seek solutions from OPM-related mechanisms. Coherent or partially coherent summation of orthogonally polarized waves~\citep[e.g.,][]{2019MNRAS.488.2018D, 2021MNRAS.501.2156D, 2023MNRAS.525..840O} is able to reproduce PA curves that are vastly deviated from RVM curves. Our model is an extension of the partial-coherence model by \citet{2023MNRAS.525..840O}. In the wave summation models, both changing the wave mode amplitude ratio (mode ratio driven) and changing the phase lag between wave modes (phase lag driven) could result in the polarization vector's rotation. We will firstly show below that the change of phase lag on different pulse longitudes could be the major reason for the polarization rotation in B1919$+$21's single pulses. Some supplementary discussions on alternative modeling and model comparisons will be presented in Section~\ref{sec:comments_on_modeling}.

        \subsection{A phenomenological model: phase lag driven}\label{sec:two-plasma}
        
\begin{figure}
		\centering
		\includegraphics[scale=0.4]{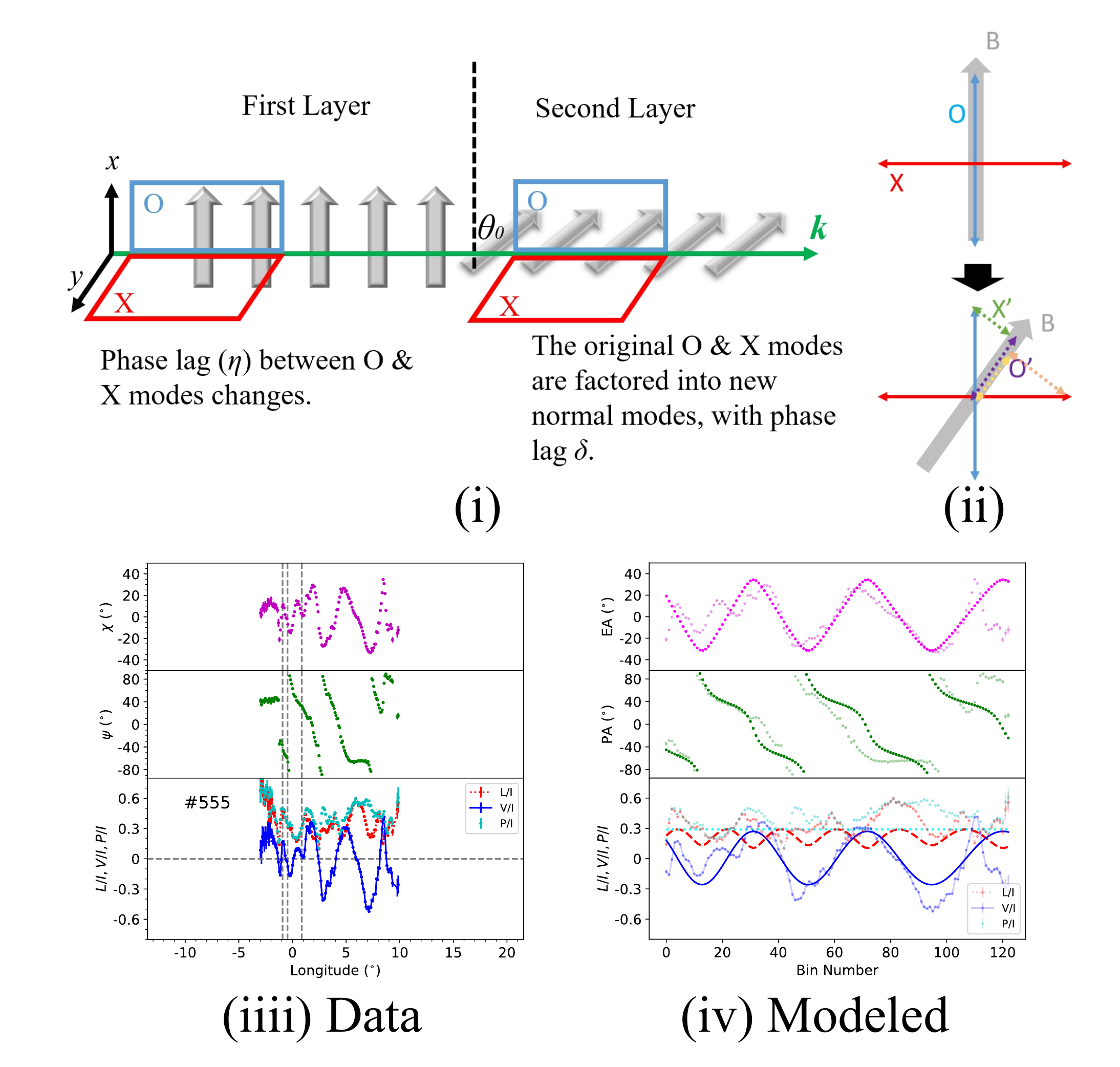}
	   \caption{(i) A schematic diagram for the phenomenological model. The gray arrows represent magnetic fields. The long green arrow with the label ``$\boldsymbol{k}$'' means the direction of wave propagation, where refraction is neglected. The blue rectangle with the label ``O'' and the red parallelogram with the label ``X'' represent O\text{-}mode wave and X\text{-}mode wave. (ii) A cross-sectional view of (i), and also a schematic diagram for the difference between two plasma layers. (iii) linear, circular, total polarization fraction, as well as PA and EA on each bin of pulse \#555 in Figure~\ref{fig:pulses}. (iv) Comparison between pulse \#555 (pale dots with error bars) and the pulse simulated through the phenomenological model, with parameters derived from fitting the model to pulse \# 555. The details are in the text. The horizontal axis is the number of sample bins, which is equivalent to the longitude. The red dashed line, the blue line, and the cyan line are linear, circular, and total polarization fractions on each bin. The green and magenta dots are PA and EA on each bin.\label{fig:modeling}}
       \end{figure}

        To reproduce the observed polarization variations in single pulses, here we put forward a phenomenological model based on propagational effects. Based on theories like those in~\cite{1979ApJ...229..348C,1998Ap&SS.262..379L,2000A&A...355.1168P}, a pulsar magnetosphere could be divided into the adiabatic walking region (the magnetic field is strong enough and normal modes are linearly polarized) and the polarization limiting region (the weakening or deflection of local magnetic fields introduces more circular polarization components). Although there is no strict boundary between the two regions, we simplify the magnetosphere into two plasma layers, where two initially mutually orthogonal radio waves (O wave and X wave) propagate. Directions of magnetic fields in two layers are fixed separately: in the first layer, the magnetic field is aligned with the O wave electric field vector; in the second layer, the magnetic field has an angle $\theta_{0}$ deflected from the magnetic field of the first layer. The phase lag between the O wave and X wave increases by a value $\eta$ through propagating in the first layer. In the second layer, the original O and X waves are decomposed into new normal wave modes, between which the phase lag increases by $\delta$ through propagating. To include depolarization, we also follow \cite{2023MNRAS.525..840O} to introduce coherence $C$ into our modeling, and then a part of the waves is added incoherently. The mathematical details of our model construction are in Appendix~\ref{sec:appendix_A}. Schematic diagrams of the model described above are shown in Figure~\ref{fig:modeling} (i).

       There are five parameters in the model: (1) initial amplitude ratio $E_{\mathrm{O}}/E_{\mathrm{X}}$; (2) phase lag produced in propagating through the first layer $\eta$; (3) magnetic field deflection angle $\theta_{0}$ (in the second layer, also shown in Figure~\ref{fig:modeling} (i)); (4) phase lag produced in propagating through the second layer $\delta$; and (5) coherence $C$. To fit a pulse with this model, we set $E_{\mathrm{O}}/E_{\mathrm{X}}$, $\theta_{0}$, $\delta$, and $C$ to constants along the pulse longitude, and set $\eta$ to a quadratic function of pulse longitude (or equally, of bin number, $\eta=a_{\eta}(\mathrm{bin}/\mathrm{bin}_{\mathrm{tot}})^2+b_{\eta}(\mathrm{bin}/\mathrm{bin}_{\mathrm{tot}})+c_{\eta}$). We apply the modeling to \#555 for example. Since our modeling does not care about absolute intensity $I$, we redraw \#555 in Figure~\ref{fig:modeling} (iii) with $V/I$, $L/I$ and $P/I$. The first 20 bins, where PA is almost horizontal, are not included in our fitting because they look like one pure mode. We fit $(Q/I, U/I, V/I)$ of \#555. The least chi-square fitting result is shown in Figure~\ref{fig:modeling} (iv). Our model could reproduce the main characteristics of the pulse, but performs badly in fitting details. Appropriate functions of $C$ and $E_{\mathrm{O}}/E_{\mathrm{X}}$ versus bin number should improve the fitting, but since the two functions are lack of physical justification at present, we will not go into detail. Anyway, the fitted parameters are: (1) $E_{\mathrm{O}}/E_{\mathrm{X}}=1.01\pm0.01$; (2) $a_{\eta}=-4.1\pm0.6$, $b_{\eta}=22.4\pm0.6$, $c_{\eta}=-2.6\pm0.2$; (3) $\theta_{0}=77^{\circ}\pm 2^{\circ}$; (4) $\delta=1.2\pm0.2$; (5) $C=0.390\pm0.006$. The result indicates that when the phase lag between normal modes rapidly varies along pulse longitudes, rapid change of polarization pattern could take place in single pulses.
       
       A natural requirement for our model to work is that none of the orthogonal modes should dominate too largely, in other words, $E_{\mathrm{O}}/E_{\mathrm{X}}$ should be close to 1 (see Figure~\ref{fig:different_paras} for some examples). Finally, we would like to emphasize that to explain the observed polarization rotations, one plasma layer, for which the normal wave modes are elliptical, is equivalent to the two plasma layers described above. The ``one plasma layer'' version of the propagational model is just the partial-coherence model by \citet{2023MNRAS.525..840O}. We introduce two plasma layers for representing different medium properties in different parts of the pulsar magnetosphere, just as we have justified at the beginning of this section.

        \subsection{Calculations on the phase lag between wave modes}\label{sec:phase_lag_calc}

        After all, the magnetosphere is much simplified in either one-plasma or two-plasma models, but we argue that the phase lag between wave modes plays a key role in the physics behind the observed rapid rotating single pulse polarization orientations, which needs further quantitative explorations.
        
        Observed single pulses' PA can monotonically rotate by values larger than $\pi$, which requires the phase lag difference, between the end longitude and the start longitude, to be greater than $2\pi$, and thus phase lags themselves should be at least on the same order of $2\pi$. This requirement may put some constraints on the magnetospheric parameters. 

\begin{figure}
		\centering
		\includegraphics[scale=0.35]{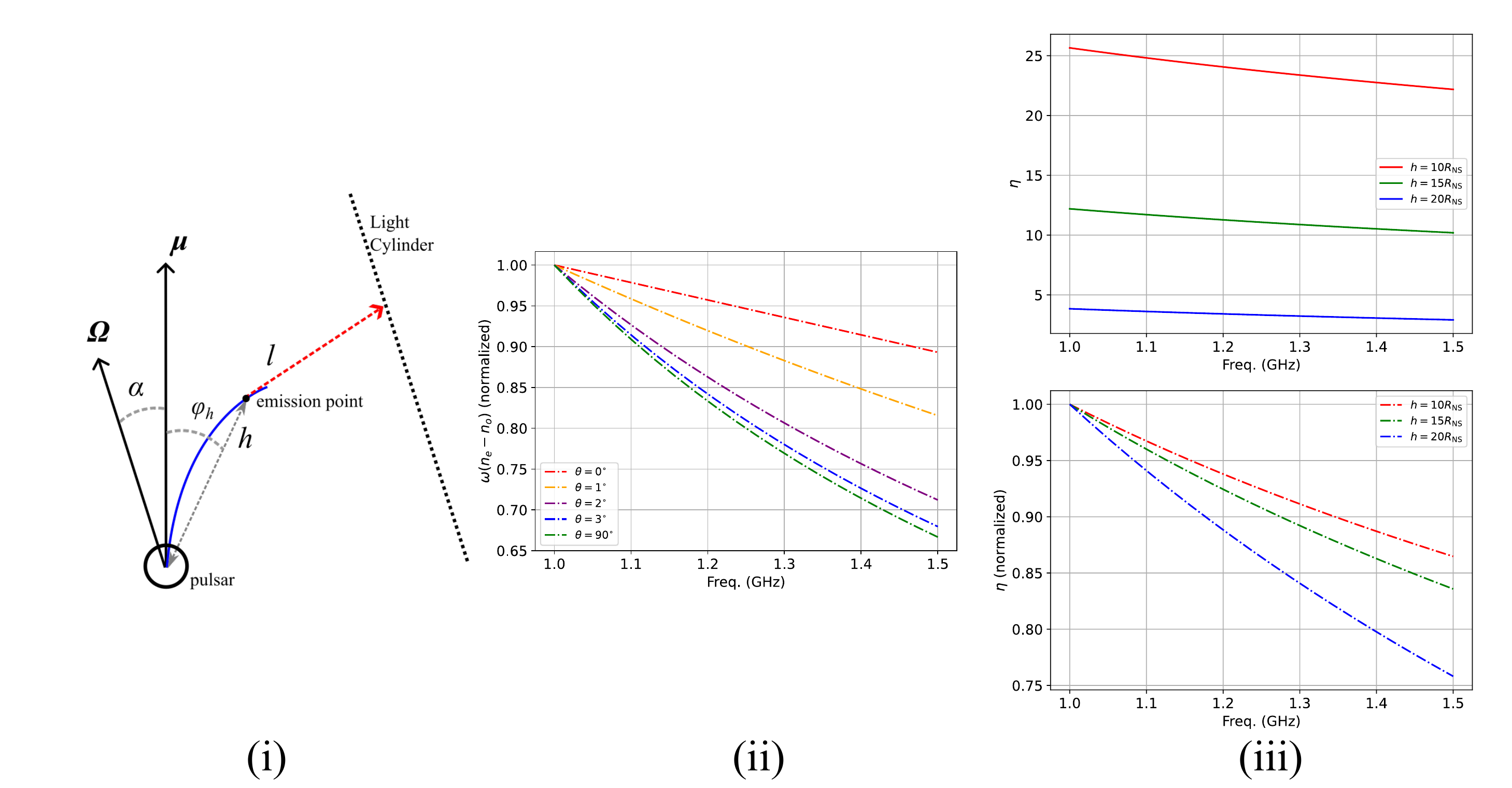}
	   \caption{(i) A schematic diagram of the phase lag calculation. (ii) Results of numerically solving Equation~\ref{eq:dispersion_rel}, presented in $\omega \cdot (n_{\mathrm{X}}-n_{\mathrm{O}})$'s dependence on frequency, given different $\theta$. For each curve, the values are normalized with the maximum of the curve, namely, the $\omega \cdot (n_{\mathrm{X}}-n_{\mathrm{O}})$ value at 1 GHz. $\gamma$ is chosen to be $100$, and $n_{e}=10^{9}\mathrm{cm}^{-3}$. (iii) Phase lags (original values and normalized values) between O and E modes versus frequency, with parameters $\kappa=10^{2}$, $\gamma=10^{3}$, $\alpha=0^{\circ}$ and $\phi_{h}=1^{\circ}$. Red, green, and blue lines represent results calculated with $h=$ 10, 15, and 20 times $R_{\mathrm{NS}}$. For details of the theoretical constructions, please refer to the text in Section~\ref{sec:phase_lag_calc}.\label{fig:phase_lag_calc}}
       \end{figure}

        Theoretical estimations of phase lag start from the dispersion relation of normal modes. We adopt the dispersion relation of O and X modes in a relativistic cold plasma with a single bulk velocity and infinitely large magnetic field~(\citealp{1977PASA....3..120M}, \citealp{1986ApJ...302..120A}), which could be represented as:

        \begin{equation}
            (1-n_{\mathrm{O}}^{2}\cos^{2}\theta)\left[1-\dfrac{\omega_{\mathrm{p}}^{2}}{\omega^{2}\gamma^{3}(1-n_{\mathrm{O}}\beta \cos\theta)^{2}}\right]-n_{\mathrm{O}}^{2}\sin^{2}\theta = 0, \quad\quad\quad n_{\mathrm{E}}=1
            \label{eq:dispersion_rel}
        \end{equation}

        \noindent where $n_{\mathrm{O}}$ and $n_{\mathrm{X}}$ are the refraction indices of O and X modes, $\theta$ is the angle between wave propagation direction and local magnetic field, and $\omega_{\mathrm{p}}=\sqrt{e^{2}n_{e}/\epsilon_{0}m_{e}}$ is the plasma (angular) frequency. We consider the electron-positron plasma here. $\gamma=\sqrt{1/(1-v^{2}/c^{2})}=\sqrt{1/(1-\beta^{2})}$ is the Lorentz factor of magnetospheric particles. Propagation on a trajectory $L$ leads to a phase lag:
        \begin{equation}
            \eta = \int_{L}(k_{\mathrm{X}}-k_{\mathrm{O}})dl = \int_{L}\dfrac{\omega}{c}(n_{\mathrm{X}}-n_{\mathrm{O}})dl
            \label{eq:phase_lag}
        \end{equation}
        \noindent where $k_{\mathrm{X}}$ and $k_{\mathrm{O}}$ refer to the wavenumbers of the X and O waves. $l$ is related to $\theta$, so $n_{\mathrm{O}}$ is the only variable depending on $l$ in the integrand in Equation~\ref{eq:phase_lag}, according to Equation~\ref{eq:dispersion_rel}. In a realistic magnetosphere, $\theta$ changes during propagation because the magnetic field's direction is not fixed along the ray trajectory. In other words, $\theta$ also depends on $l$. Considering the fact that observed rapid rotating polarization orientations in single pulses have a very weak dependence on frequency, $n_{\mathrm{O}}$'s frequency dependence, which is actually affected by the $\theta$ value, is to be investigated below. The numerical solution of Equation~\ref{eq:dispersion_rel} shown in Figure~\ref{fig:phase_lag_calc} (ii) tells us that the frequency dependence of $n_{\mathrm{O}}$ (or $\omega(n_{\mathrm{X}}-n_{\mathrm{O}})$) is weakest at $\theta=0^{\circ}$ and strongest at $\theta=90^{\circ}$. The case for very small $\theta$ works better for explaining the weak frequency dependence of rapid polarization rotations that we observe. 
        
        Given the magnetic field configuration and the distribution of plasma particles, the phase lag between O and X waves of every ray trajectory could be derived. In a dipolar magnetic field, $\theta$ always becomes larger when waves propagate outward, so a small $\theta$ in the whole propagational trajectory prefers the emission to originate relatively close to the magnetic axis, which is also in accord with the small $\beta$ value we have mentioned in Section~\ref{sec:PA_geo}. For now, we only consider the aligned rotator ($\alpha=0^{\circ}$). For oblique rotators, discussions are made in Section~\ref{sec:comments_on_modeling}.

        Our main assumptions and approximations for the calculation are (a) pure dipolar magnetic field; (b) ray trajectories are straight lines, where refraction is neglected; (c) wave propagation direction follows only the tangential direction of magnetic field at the emission point; and (d) the plasma particle number density decreases as the distance to pulsar center $r$ increases, in the form (similar to Equation (51) in~\cite{1975ApJ...196...51R}):

        \begin{equation}
            n_{e}(r) = \kappa \cdot n_{\mathrm{GJ,surf}}\cdot \left(\dfrac{R_{\mathrm{NS}}}{r}\right)^{3} =\kappa \cdot 7\times 10^{10}\left(\dfrac{B_{\mathrm{surf}}}{10^{12}\mathrm{G}}\right)\left(\dfrac{P}{1\mathrm{s}}\right)^{-1}\cdot \left(\dfrac{R_{\mathrm{NS}}}{r}\right)^{3} \mathrm{cm^{-3}}
            \label{eq:number_density}
        \end{equation}

        \noindent where $n_{\mathrm{GJ,surf}}$ is the Goldreich-Julian number density at pulsar surface (Equation (9) in~\cite{1969ApJ...157..869G}), $B_{\mathrm{surf}}$ is the surface magnetic field strength estimated with period $P$ and period derivative $\dot{P}$ (see \citealp{2022ARA&A..60..495P} for example). The neutron star radius $R_{\mathrm{NS}}$ is assumed simply to be 10km.

        The integration of Equation~\ref{eq:phase_lag} begins at the emission point and ends at the light cylinder. Given $l$, $n_{\mathrm{O}}$ is determined through calculating $\theta$---then the whole integrand is determined. A schematic diagram of the integration path is shown in Figure~\ref{fig:phase_lag_calc} (i). With $\kappa=100$, $\gamma=1000$, the phase lags calculated at different frequencies are plotted in Figure~\ref{fig:phase_lag_calc} (iii). Under these parameter values, $\eta$s lie on the same orders of $2\pi\sim 6\pi$. A larger $\kappa$ or a smaller $\gamma$ cause the phase lag to be larger. When emission height is lower, the phase lag is also larger, and the frequency dependence is weaker. The absence of spectral evolution of observed polarization rotations suggests a small emission height for B1919$+$21.

  \section{Discussion}\label{sec:discussion}
        \subsection{Comments on modeling in Section~\ref{sec:modeling}}\label{sec:comments_on_modeling}

        Some of our choices of parameters in Section~\ref{sec:two-plasma} are mainly made for simplicity. We let $\eta$ vary but keep $\delta$ constant---this assumption could be adjusted to let both $\eta$ and $\delta$ change, which could happen in real magnetospheres because of the difference in plasma particle distribution, for example. In addition, $E_{\mathrm{O}}/E_{\mathrm{X}}$, $\theta_{0}$ and $C$ need not to be constant along pulse longitudes. In observed pulses, the $P/I$ curve seems to be related to the $I$ profile, which should be associated with coherence $C$. $C$ could also be different in two plasma layers. Some simulated pulses with different parameters, including one with varying $C$, are shown in Figure~\ref{fig:different_paras} in Appendix~\ref{sec:appendix_B}. Our phenomenological model could be applied to explain other polarization behaviors in pulsars, but we do not go into detail here.

        Equation~\ref{eq:dispersion_rel} is based on the assumption that the magnetic field is large enough. Quantitatively, validation of Equation~\ref{eq:dispersion_rel} requires $\omega_{B}\gg\omega$, where $\omega_{B}=eB/m_{e}c$. With dipole field $B\approx10^{12}(R_{\mathrm{NS}}/r)^{3}\mathrm{G}$ we have $\omega_{B}=1.8\times10^{19}(R_{\mathrm{NS}}/r)^{3}\mathrm{s}^{-1}$. So, Equation~\ref{eq:dispersion_rel} is valid at least within $10^{3}R_{\mathrm{NS}}$ from the neutron star centroid. And because plasma particle number density decreases quickly when $r$ increases (Equation~\ref{eq:number_density}), the integration of Equation~\ref{eq:phase_lag} will hardly differ between integrating to the light cylinder and integrating to $\sim 10^{3}R_{\mathrm{NS}}$. The inner magnetosphere contributes most to the phase lag---this is also valid for oblique rotators ($\alpha>0$).

        $\eta$ in Section~\ref{sec:phase_lag_calc} is evaluated to be on the order of $2\pi$. To achieve rapid polarization rotations in single pulses, $\eta$ might be larger, which demands that the corresponding $\kappa$ to be larger or $\gamma$ to be smaller. But the requirement of low emission height should always be satisfied to achieve a small $\theta$ in Equation~\ref{eq:dispersion_rel}.

        The phase\text{-}lag\text{-}driven model that we present performs well in understanding large rotations of PA, but the requirement of nearly equal mode strength ratio remains a question (see Figure~\ref{fig:different_paras} for a simulated pulse with $E_{\mathrm{O}}/E_{\mathrm{X}}=1.4$, where the monotonic rotation of PA ceases). We could also consider modeling that is based on the rotation of the emitted polarization orientation or on the variable mode amplitude ratio. The pure RVM-type PA curve can only cover $180^{\circ}$ of PA values. The change of wave modes amplitude ratio could also lead to rotations of polarization orientation, but if we keep the phase lag constant and consider only one pair of OPM, the PA variation will not be greater than $180^{\circ}$. After taking RVM into account, amplitude ratio-driven PA variation could cover $360^{\circ}$. This PA variation is still inadequate compared with some very large PA rotations (e.g. \#2565 in Figure~\ref{fig:pulses}) of our observed single pulses. In order to achieve PA rotation over $360^{\circ}$, the emitted radio wave's polarization orientation should initially be rotating at different longitudes before propagating, or there should be more pairs of OPMs added coherently. A separate paper is in preparation to discuss this.

        The requirement of low emission height could be related to the possible small impact angle of viewing B1919$+$21. But B1919$+$21 may not be ``special'' for its rapid rotating polarization orientations in single pulses because many pulsars actually lack detailed single\text{-}pulse investigations of polarization. A more refined model might be achieved after more similar or relevant phenomena are analyzed.
        
		\subsection{On asymmetries in B1919$+$21's polarization patterns}\label{sec:asymmetry}

        We would like to emphasize two asymmetries in our observational facts. One is that from the RVM fitting, the centroid of the PA curve seems to arrive manifestly earlier than the centroid (valley) of the integrated pulse's profile of total intensity, which contradicts the model of relativistic aberration (Blaskiewicz-Cordes-Wasserman (BCW) model, \citealt{1991ApJ...370..643B}), assuming a symmetrically filled beam. The other, which is more rigid, is that most monotonic long PA curves (cover $>180^{\circ}$) in single pulses have a negative slope, while only very few of them have a positive slope. Both asymmetries could be interpreted as consequences of the asymmetric distribution of magnetospheric plasma particles' number density or Lorentz factor, or of discharge positions near the pulsar surface. In the case of our theoretical explanation, PA slopes imply the increase or decrease of phase lag between O and X modes versus longitude, which depends on the distributions mentioned above.

        Considering the fact that PA curves could change greatly between adjacent pulses (see \#274 and \#275 in Figure~\ref{fig:pulses} for examples), and that the dispersion relation (Equation~\ref{eq:dispersion_rel}) depends sensitively on the Lorentz factor $\gamma$, the polarization pattern of single pulses may mostly be affected by the $\gamma$ distribution of plasma particles in the magnetosphere. Distributions of particle number density and discharge positions seem more likely to be almost fixed, although they may be asymmetrical about the fiducial plane (where line of sight, magnetic axis, and rotational axis are coplanar), probably due to some particular surface structures~\citep[e.g.,][]{2023-QCS19,2024AN....34540010W}.


        

		\section{Conclusion}\label{sec:conclusion}
		
		The FAST observational results of PSR~B1919$+$21 are presented in this manuscript. The complex PA curve and circular polarization variation of the integrated profile result from the rapid rotations of polarization orientations in single pulses. In many observed single pulses, the PA could monotonically rotate over $180^{\circ}$ or even $360^{\circ}$ across longitudes, and is always accompanied by circular polarization variations that could change handedness. Most of those quasi-monotonic PA curves have negative slopes. 
        
        We attribute the phenomena to the difference in normal modes' phase lags between different pulse longitudes, and build a simple model based on propagational effects that can reproduce the observed polarization patterns in single pulses. An appropriate value of phase lag between O and E waves requires appropriate values of pulsar magnetosphere's multiplicity $\kappa$ and secondary plasma's Lorentz factor $\gamma$, which could be related to magnetosphere compositions. The emission height should be low enough to fit the weak frequency dependence of the polarization rotations. We argue that the distribution of plasma number density, of discharge points near the pulsar surface, or of secondary particles' Lorentz factor $\gamma$, should be asymmetric around the fiducial plane to explain the asymmetry observed in single pulses' monotonic PA curves' slopes. Further explorations into particle acceleration processes and pulsar surface properties may help further our understanding of them.
        
        Still, we are not sure if B1919$+$21 is ``special'' for exhibiting such a single\text{-}pulse polarization pattern. Except for the requirement of more single\text{-}pulse investigation of this pulsar under other frequency bands, we appeal that researchers do more single\text{-}pulses studies on more known pulsars for a better understanding of radio pulsars' radiation and magnetospheres.

		\acknowledgments
		
		S.S.C. thanks Prof. Gregory Herczeg at KIAA, Peking University. Through his ``Advanced Writing for Astronomy'' course, S.S.C.'s writing of this paper was improved. Shenglan Sun at Peking University also gave useful comments on the paper during the writing course by Prof. Gregory Herczeg. An anonymous reviewer is acknowledged for providing us with many suggestions. J.D. acknowledges funding by the National Science Centre, Poland, grant no. 2023/49/B/ST9/01783. All data used in this work is from the FAST (Five-hundred-meter Aperture Spherical radio Telescope) (https://cstr.cn/31116.02.FAST). FAST is a Chinese national mega-science facility, operated by National Astronomical Observatories, Chinese Academy of Sciences. This work is supported by the National
		SKA Program of China (2020SKA0120100), the National Natural Science Foundation of China (Nos. 12003047 and 12133003), and the Strategic Priority Research Program of the Chinese Academy of Sciences (No. XDB0550300).

		%

		
        \appendix
        \section{Formula for constructing the phenomenological model}\label{sec:appendix_A}

        The mathematical details of the model described in Section~\ref{sec:two-plasma} are presented in this section. We mark the initial O and X waves' electric field vector amplitudes as $E_{\mathrm{O}}$ and $E_{\mathrm{X}}$. $E^{(C)}_{\mathrm{O},0}=C\cdot E_{\mathrm{O}}$ and $E^{(C)}_{\mathrm{X},0}=C\cdot E_{\mathrm{X}}$ represent the coherently added part, while $E^{(I1)}_{\mathrm{O},0}=(1-C)\cdot E_{\mathrm{O}}$, $E^{(I1)}_{\mathrm{X},0}=0$ and $E^{(I2)}_{\mathrm{O},0}=0$, $E^{(I2)}_{\mathrm{X},0}=(1-C)\cdot E_{\mathrm{X}}$ represent the incoherently added part (same as the settings in~\citealp{2023MNRAS.525..840O}). After propagation in the first plasma layer, the electric field vectors become
        \begin{flalign}
        & E^{(C;I)}_{\mathrm{O},1} = E^{(C;I)}_{\mathrm{O},0}\\
        & E^{(C;I)}_{\mathrm{X},1} = E^{(C;I)}_{\mathrm{X},0}e^{i\eta}
        \end{flalign}
        and after propagating through the second plasma layer, the electric field vectors become
        \begin{flalign}
        & E^{(C;I)}_{\mathrm{O},2} = E^{(C;I)}_{\mathrm{O},1}\cos{\theta_{0}}+E^{(C;I)}_{\mathrm{X},1}\sin{\theta_{0}}\\
        & E^{(C;I)}_{\mathrm{X},2} = (-E^{(C;I)}_{\mathrm{O},1}\sin{\theta_{0}}+E^{(C;I)}_{\mathrm{X},1}\cos{\theta_{0}})e^{i\delta}
        \end{flalign}
        Then, we return to the original coordinates, where the electric field vectors in the original O and X wave directions ($x$ and $y$ in Figure~\ref{fig:modeling}) are

        \begin{flalign}
            & E^{(C;I)}_{x} = E^{(C;I)}_{\mathrm{O},2}\cos{\theta_{0}}-E^{(C;I)}_{\mathrm{X},2}\sin{\theta_{0}} \\
            & E^{(C;I)}_{y} = E^{(C;I)}_{\mathrm{O},2}\sin{\theta_{0}}+E^{(C;I)}_{\mathrm{X},2}\cos{\theta_{0}}
        \end{flalign}

        \noindent Finally, we can calculate Stokes parameters~\citep[e.g.,][]{1998clel.book.....J}:

        \begin{flalign}
            & I^{(C;I)} = E^{(C;I)}_{x}(E^{(C;I)}_{x})^{*}+E^{(C;I)}_{y}(E^{(C;I)}_{y})^{*}\\
            & Q^{(C;I)} = E^{(C;I)}_{x}(E^{(C;I)}_{x})^{*}-E^{(C;I)}_{y}(E^{(C;I)}_{y})^{*}\\
            & U^{(C;I)} = 2\mathrm{Re}[(E^{(C;I)}_{x})^{*}E^{(C;I)}_{y}]\\
            & V^{(C;I)} = 2\mathrm{Im}[(E^{(C;I)}_{x})^{*}E^{(C;I)}_{y}]
        \end{flalign}

        \noindent The total Stokes parameters are

        \begin{equation}
            S = S^{(C)} + S^{(I1)} + S^{(I2)}
        \end{equation}

        \noindent where $S=(I,Q,U,V)$. With Stokes parameters, we can calculate PA and EA

        \begin{flalign}
            & \mathrm{PA}=0.5\arctan (U/Q)\label{eq:PA}\\
            & \mathrm{EA} = 0.5\arctan(V/\sqrt{Q^{2}+U^{2}})\label{eq:EA}
        \end{flalign}
        
        \noindent and draw the simulated pulse in Figure~\ref{fig:modeling}.

        \section{Six simulated pulses}\label{sec:appendix_B}
\begin{figure}
		\centering
		\includegraphics[scale=0.4]{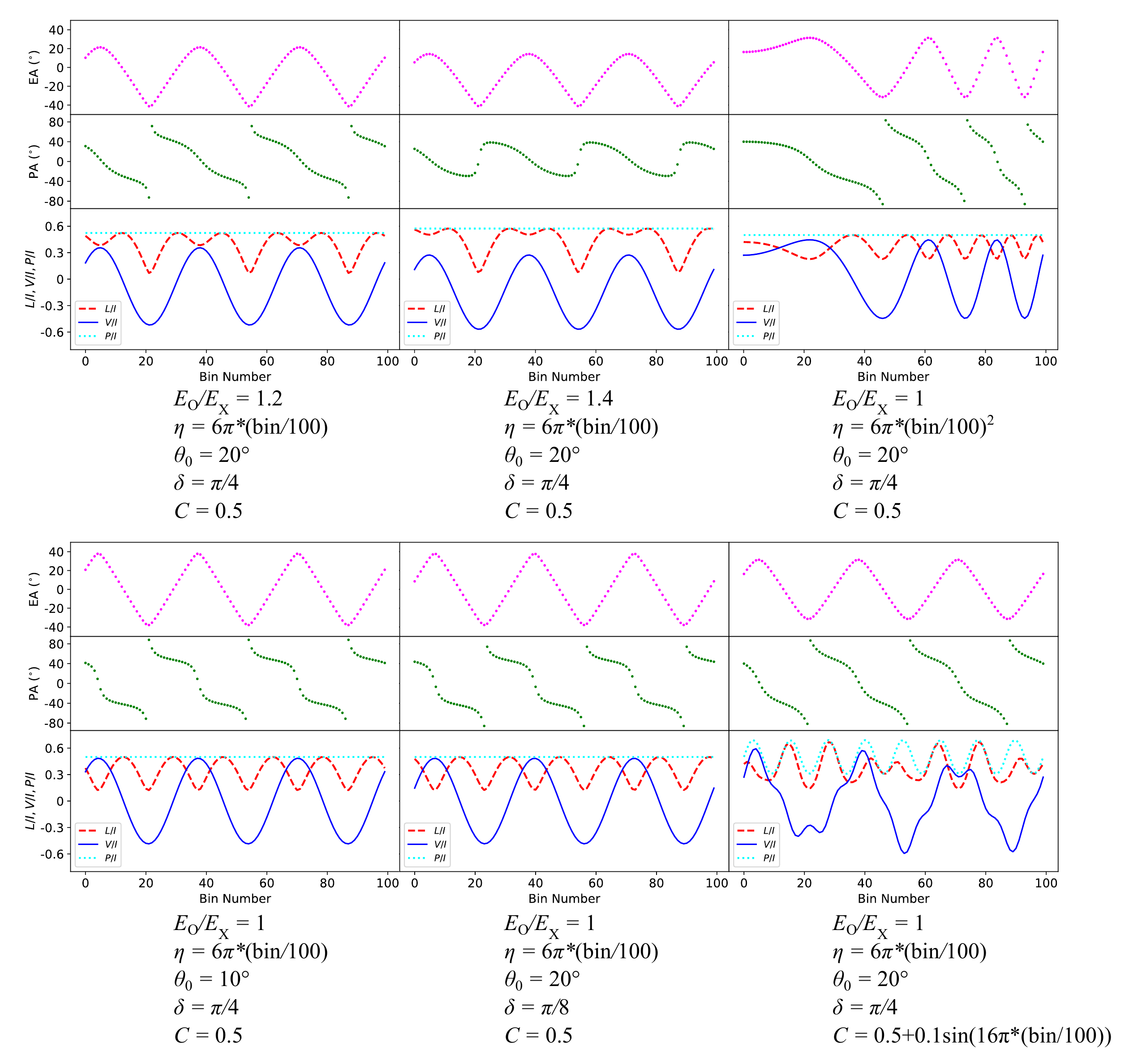}
	   \caption{Six pulses simulated from the phenomenological model. The meanings of the dots and lines are the same as those in Figure~\ref{fig:modeling} (iv). The parameters used to simulate the pulses are given below each plot.\label{fig:different_paras}}
        \end{figure}

        Six simulated pulses with different groups of parameters are shown in Figure~\ref{fig:different_paras}.

\begin{thebibliography}{}
				\expandafter\ifx\csname natexlab\endcsname\relax\def\natexlab#1{#1}\fi
				\providecommand{\url}[1]{\href{#1}{#1}}
				
				\bibitem[{{Arons} \& {Barnard}(1986)}]{1986ApJ...302..120A}
				{Arons}, J., \& {Barnard}, J.~J. 1986, \apj, 302, 120
				
				\bibitem[{{Barnard} \& {Arons}(1986)}]{1986ApJ...302..138B}
				{Barnard}, J.~J., \& {Arons}, J. 1986, \apj, 302, 138
				
				\bibitem[{{Bera} {et~al.}(2024){Bera}, {James}, {McKinnon}, {Ekers}, {Dial}, {Deller}, {Bannister}, {Glowacki}, \& {Shannon}}]{2024arXiv241114784B}
				{Bera}, A., {James}, C.~W., {McKinnon}, M.~M., {et~al.} 2024, arXiv e-prints, arXiv:2411.14784
				
				\bibitem[{{Beskin} \& {Philippov}(2012)}]{2012MNRAS.425..814B}
				{Beskin}, V.~S., \& {Philippov}, A.~A. 2012, \mnras, 425, 814
				
				\bibitem[{{Blaskiewicz} {et~al.}(1991){Blaskiewicz}, {Cordes}, \& {Wasserman}}]{1991ApJ...370..643B}
				{Blaskiewicz}, M., {Cordes}, J.~M., \& {Wasserman}, I. 1991, \apj, 370, 643
				
				\bibitem[{{Cao} {et~al.}(2024){Cao}, {Jiang}, {Dyks}, {Hao}, {Lee}, {Li}, {Lu}, {Pan}, {Wang}, {Wang}, {Xu}, {Xu}, \& {Xu}}]{2024ApJ...973...56C}
				{Cao}, S., {Jiang}, J., {Dyks}, J., {et~al.} 2024, \apj, 973, 56
				
				\bibitem[{{Cheng} \& {Ruderman}(1979)}]{1979ApJ...229..348C}
				{Cheng}, A.~F., \& {Ruderman}, M.~A. 1979, \apj, 229, 348
				
				\bibitem[{{Cordes}(1975)}]{1975ApJ...195..193C}
				{Cordes}, J.~M. 1975, \apj, 195, 193
				
				\bibitem[{{Dyks}(2017)}]{2017MNRAS.472.4598D}
				{Dyks}, J. 2017, \mnras, 472, 4598
				
				\bibitem[{{Dyks}(2019)}]{2019MNRAS.488.2018D}
				---. 2019, \mnras, 488, 2018
				
				\bibitem[{{Dyks}(2023)}]{2023MNRAS.522.1480D}
				---. 2023, \mnras, 522, 1480
				
				\bibitem[{{Dyks} {et~al.}(2010){Dyks}, {Rudak}, \& {Demorest}}]{2010MNRAS.401.1781D}
				{Dyks}, J., {Rudak}, B., \& {Demorest}, P. 2010, \mnras, 401, 1781
				
				\bibitem[{{Dyks} {et~al.}(2021){Dyks}, {Weltevrede}, \& {Ilie}}]{2021MNRAS.501.2156D}
				{Dyks}, J., {Weltevrede}, P., \& {Ilie}, C. 2021, \mnras, 501, 2156
				
				\bibitem[{{Everett} \& {Weisberg}(2001)}]{2001ApJ...553..341E}
				{Everett}, J.~E., \& {Weisberg}, J.~M. 2001, \apj, 553, 341
				
				\bibitem[{{Goldreich} \& {Julian}(1969)}]{1969ApJ...157..869G}
				{Goldreich}, P., \& {Julian}, W.~H. 1969, \apj, 157, 869
				
				\bibitem[{{Hassall} {et~al.}(2012){Hassall}, {Stappers}, {Hessels}, {Kramer}, {Alexov}, {Anderson}, {Coenen}, {Karastergiou}, {Keane}, {Kondratiev}, {Lazaridis}, {van Leeuwen}, {Noutsos}, {Serylak}, {Sobey}, {Verbiest}, {Weltevrede}, {Zagkouris}, {Fender}, {Wijers}, {B{\"a}hren}, {Bell}, {Broderick}, {Corbel}, {Daw}, {Dhillon}, {Eisl{\"o}ffel}, {Falcke}, {Grie{\ss}meier}, {Jonker}, {Law}, {Markoff}, {Miller-Jones}, {Osten}, {Rol}, {Scaife}, {Scheers}, {Schellart}, {Spreeuw}, {Swinbank}, {ter Veen}, {Wise}, {Wijnands}, {Wucknitz}, {Zarka}, {Asgekar}, {Bell}, {Bentum}, {Bernardi}, {Best}, {Bonafede}, {Boonstra}, {Brentjens}, {Brouw}, {Br{\"u}ggen}, {Butcher}, {Ciardi}, {Garrett}, {Gerbers}, {Gunst}, {van Haarlem}, {Heald}, {Hoeft}, {Holties}, {de Jong}, {Koopmans}, {Kuniyoshi}, {Kuper}, {Loose}, {Maat}, {Masters}, {McKean}, {Meulman}, {Mevius}, {Munk}, {Noordam}, {Orr{\'u}}, {Paas}, {Pandey-Pommier}, {Pandey}, {Pizzo}, {Polatidis}, {Reich}, {R{\"o}ttgering}, {Sluman}, {Steinmetz}, {Sterks}, {Tagger}, {Tang},
					{Tasse}, {Vermeulen}, {van Weeren}, {Wijnholds}, \& {Yatawatta}}]{2012A&A...543A..66H}
				{Hassall}, T.~E., {Stappers}, B.~W., {Hessels}, J.~W.~T., {et~al.} 2012, \aap, 543, A66
				
				\bibitem[{{Hewish} {et~al.}(1968){Hewish}, {Bell}, {Pilkington}, {Scott}, \& {Collins}}]{1968Natur.217..709H}
				{Hewish}, A., {Bell}, S.~J., {Pilkington}, J.~D.~H., {Scott}, P.~F., \& {Collins}, R.~A. 1968, \nat, 217, 709
				
				\bibitem[{{Hobbs} {et~al.}(2006){Hobbs}, {Edwards}, \& {Manchester}}]{2006MNRAS.369..655H}
				{Hobbs}, G.~B., {Edwards}, R.~T., \& {Manchester}, R.~N. 2006, \mnras, 369, 655
				
				\bibitem[{{Hotan} {et~al.}(2004){Hotan}, {van Straten}, \& {Manchester}}]{2004PASA...21..302H}
				{Hotan}, A.~W., {van Straten}, W., \& {Manchester}, R.~N. 2004, \pasa, 21, 302
				
				\bibitem[{{Jackson}(1998)}]{1998clel.book.....J}
				{Jackson}, J.~D. 1998, {Classical Electrodynamics, 3rd Edition} (Wiley-VCH \& Higher Education Press, Beijing (reprinted in 2013)), 301
				
				\bibitem[{{Jiang} {et~al.}(2022){Jiang}, {Wang}, {Xu}, {Xu}, {Zhang}, {Wang}, {Zhou}, {Zhang}, {Niu}, {Lee}, {Zhang}, {Han}, {Li}, {Zhu}, {Dai}, {Feng}, {Jing}, {Li}, {Luo}, {Miao}, {Niu}, {Tsai}, {Wang}, {Wang}, {Xu}, {Yang}, {Yang}, {Yao}, \& {Yuan}}]{2022RAA....22l4003J}
				{Jiang}, J.-C., {Wang}, W.-Y., {Xu}, H., {et~al.} 2022, Research in Astronomy and Astrophysics, 22, 124003
				
				\bibitem[{{Jiang} {et~al.}(2020){Jiang}, {Tang}, {Hou}, {Liu}, {Kr{\v{c}}o}, {Qian}, {Sun}, {Ching}, {Liu}, {Duan}, {Yue}, {Gan}, {Yao}, {Li}, {Pan}, {Yu}, {Liu}, {Li}, {Peng}, {Yan}, \& {FAST Collaboration}}]{2020RAA....20...64J}
				{Jiang}, P., {Tang}, N.-Y., {Hou}, L.-G., {et~al.} 2020, Research in Astronomy and Astrophysics, 20, 064
				
				\bibitem[{{Johnston} {et~al.}(2024){Johnston}, {Mitra}, {Keith}, {Oswald}, \& {Karastergiou}}]{2024MNRAS.530.4839J}
				{Johnston}, S., {Mitra}, D., {Keith}, M.~J., {Oswald}, L.~S., \& {Karastergiou}, A. 2024, \mnras, 530, 4839
				
				\bibitem[{{Kuz'min} \& {Wu}(1992)}]{1992Ap&SS.190..209K}
				{Kuz'min}, A.~D., \& {Wu}, X. 1992, \apss, 190, 209
				
				\bibitem[{{Lower} {et~al.}(2024{\natexlab{a}}){Lower}, {Kramer}, {Johnston}, {Breton}, {Wex}, {Bailes}, {Buchner}, {Camilo}, {Oswald}, {Reardon}, {Shannon}, {Serylak}, \& {Krishnan}}]{2024MNRAS.534.3936L}
				{Lower}, M.~E., {Kramer}, M., {Johnston}, S., {et~al.} 2024{\natexlab{a}}, \mnras, 534, 3936
				
				\bibitem[{{Lower} {et~al.}(2024{\natexlab{b}}){Lower}, {Johnston}, {Lyutikov}, {Melrose}, {Shannon}, {Weltevrede}, {Caleb}, {Camilo}, {Cameron}, {Dai}, {Hobbs}, {Li}, {Rajwade}, {Reynolds}, {Sarkissian}, \& {Stappers}}]{2024NatAs...8..606L}
				{Lower}, M.~E., {Johnston}, S., {Lyutikov}, M., {et~al.} 2024{\natexlab{b}}, Nature Astronomy, 8, 606
				
				\bibitem[{{Lyubarskii} \& {Petrova}(1998)}]{1998Ap&SS.262..379L}
				{Lyubarskii}, Y.~E., \& {Petrova}, S.~A. 1998, \apss, 262, 379
				
				\bibitem[{{Lyutikov} \& {Thompson}(2005)}]{2005ApJ...634.1223L}
				{Lyutikov}, M., \& {Thompson}, C. 2005, \apj, 634, 1223
				
				\bibitem[{{Manchester} {et~al.}(2005){Manchester}, {Hobbs}, {Teoh}, \& {Hobbs}}]{2005AJ....129.1993M}
				{Manchester}, R.~N., {Hobbs}, G.~B., {Teoh}, A., \& {Hobbs}, M. 2005, \aj, 129, 1993
				
				\bibitem[{{Manchester} {et~al.}(1975){Manchester}, {Taylor}, \& {Huguenin}}]{1975ApJ...196...83M}
				{Manchester}, R.~N., {Taylor}, J.~H., \& {Huguenin}, G.~R. 1975, \apj, 196, 83
				
				\bibitem[{{Melrose} \& {Stoneham}(1977)}]{1977PASA....3..120M}
				{Melrose}, D.~B., \& {Stoneham}, R.~J. 1977, \pasa, 3, 120
				
				\bibitem[{{Mitra} {et~al.}(2015){Mitra}, {Arjunwadkar}, \& {Rankin}}]{2015ApJ...806..236M}
				{Mitra}, D., {Arjunwadkar}, M., \& {Rankin}, J.~M. 2015, \apj, 806, 236
				
				\bibitem[{{Mitra} {et~al.}(2023){Mitra}, {Melikidze}, \& {Basu}}]{2023MNRAS.521L..34M}
				{Mitra}, D., {Melikidze}, G.~I., \& {Basu}, R. 2023, \mnras, 521, L34
				
				\bibitem[{{Mitra} \& {Rankin}(2002)}]{2002ApJ...577..322M}
				{Mitra}, D., \& {Rankin}, J.~M. 2002, \apj, 577, 322
				
				\bibitem[{{Noordhuis} {et~al.}(2023){Noordhuis}, {Prabhu}, {Witte}, {Chen}, {Cruz}, \& {Weniger}}]{2023PhRvL.131k1004N}
				{Noordhuis}, D., {Prabhu}, A., {Witte}, S.~J., {et~al.} 2023, \prl, 131, 111004
				
				\bibitem[{{Olszanski} {et~al.}(2019){Olszanski}, {Mitra}, \& {Rankin}}]{2019MNRAS.489.1543O}
				{Olszanski}, T. E.~E., {Mitra}, D., \& {Rankin}, J.~M. 2019, \mnras, 489, 1543
				
				\bibitem[{{Oswald} {et~al.}(2023){Oswald}, {Karastergiou}, \& {Johnston}}]{2023MNRAS.525..840O}
				{Oswald}, L.~S., {Karastergiou}, A., \& {Johnston}, S. 2023, \mnras, 525, 840
				
				\bibitem[{{Petrova} \& {Lyubarskii}(2000)}]{2000A&A...355.1168P}
				{Petrova}, S.~A., \& {Lyubarskii}, Y.~E. 2000, \aap, 355, 1168
				
				\bibitem[{{Philippov} \& {Kramer}(2022)}]{2022ARA&A..60..495P}
				{Philippov}, A., \& {Kramer}, M. 2022, \araa, 60, 495
				
				\bibitem[{{Pilia} {et~al.}(2016){Pilia}, {Hessels}, {Stappers}, {Kondratiev}, {Kramer}, {van Leeuwen}, {Weltevrede}, {Lyne}, {Zagkouris}, {Hassall}, {Bilous}, {Breton}, {Falcke}, {Grie{\ss}meier}, {Keane}, {Karastergiou}, {Kuniyoshi}, {Noutsos}, {Os{\l}owski}, {Serylak}, {Sobey}, {ter Veen}, {Alexov}, {Anderson}, {Asgekar}, {Avruch}, {Bell}, {Bentum}, {Bernardi}, {B{\^\i}rzan}, {Bonafede}, {Breitling}, {Broderick}, {Br{\"u}ggen}, {Ciardi}, {Corbel}, {de Geus}, {de Jong}, {Deller}, {Duscha}, {Eisl{\"o}ffel}, {Fallows}, {Fender}, {Ferrari}, {Frieswijk}, {Garrett}, {Gunst}, {Hamaker}, {Heald}, {Horneffer}, {Jonker}, {Juette}, {Kuper}, {Maat}, {Mann}, {Markoff}, {McFadden}, {McKay-Bukowski}, {Miller-Jones}, {Nelles}, {Paas}, {Pandey-Pommier}, {Pietka}, {Pizzo}, {Polatidis}, {Reich}, {R{\"o}ttgering}, {Rowlinson}, {Schwarz}, {Smirnov}, {Steinmetz}, {Stewart}, {Swinbank}, {Tagger}, {Tang}, {Tasse}, {Thoudam}, {Toribio}, {van der Horst}, {Vermeulen}, {Vocks}, {van Weeren}, {Wijers}, {Wijnands}, {Wijnholds},
					{Wucknitz}, \& {Zarka}}]{2016A&A...586A..92P}
				{Pilia}, M., {Hessels}, J.~W.~T., {Stappers}, B.~W., {et~al.} 2016, \aap, 586, A92
				
				\bibitem[{{Primak} {et~al.}(2022){Primak}, {Tiburzi}, {van Straten}, {Dyks}, \& {Gulyaev}}]{2022A&A...657A..34P}
				{Primak}, N., {Tiburzi}, C., {van Straten}, W., {Dyks}, J., \& {Gulyaev}, S. 2022, \aap, 657, A34
				
				\bibitem[{{Proszynski} \& {Wolszczan}(1986)}]{1986ApJ...307..540P}
				{Proszynski}, M., \& {Wolszczan}, A. 1986, \apj, 307, 540
				
				\bibitem[{{Radhakrishnan} \& {Cooke}(1969)}]{1969ApL.....3..225R}
				{Radhakrishnan}, V., \& {Cooke}, D.~J. 1969, \aplett, 3, 225
				
				\bibitem[{{Ruderman} \& {Sutherland}(1975)}]{1975ApJ...196...51R}
				{Ruderman}, M.~A., \& {Sutherland}, P.~G. 1975, \apj, 196, 51
				
				\bibitem[{{van Straten} \& {Bailes}(2011)}]{2011PASA...28....1V}
				{van Straten}, W., \& {Bailes}, M. 2011, \pasa, 28, 1
				
				\bibitem[{{van Straten} {et~al.}(2010){van Straten}, {Manchester}, {Johnston}, \& {Reynolds}}]{2010PASA...27..104V}
				{van Straten}, W., {Manchester}, R.~N., {Johnston}, S., \& {Reynolds}, J.~E. 2010, \pasa, 27, 104
				
				\bibitem[{{Wang} {et~al.}(2024){Wang}, {Lu}, {Jiang}, {Cao}, {Wang}, {Liang}, \& {Xu}}]{2024AN....34540010W}
				{Wang}, Z., {Lu}, J., {Jiang}, J., {et~al.} 2024, Astronomische Nachrichten, 345, e20240010
				
				\bibitem[{{Weisberg} {et~al.}(1999){Weisberg}, {Cordes}, {Lundgren}, {Dawson}, {Despotes}, {Morgan}, {Weitz}, {Zink}, \& {Backer}}]{1999ApJS..121..171W}
				{Weisberg}, J.~M., {Cordes}, J.~M., {Lundgren}, S.~C., {et~al.} 1999, \apjs, 121, 171
				
				\bibitem[{Xu(2024)}]{2023-QCS19}
				Xu, R. 2024, Nuclear Physics Review, 41, 863.
				\newblock \url{http://www.npr.ac.cn/en/article/doi/10.11804/NuclPhysRev.41.QCS2023.19}
				
				\bibitem[{{Xue} {et~al.}(2023){Xue}, {Lee}, {Gao}, \& {Xu}}]{2023PhRvD.108h3009X}
				{Xue}, Z.~H., {Lee}, K.~J., {Gao}, X.~D., \& {Xu}, R.~X. 2023, \prd, 108, 083009
				
			\end{thebibliography}
		\end{document}